# Eigenstates for billiards of arbitrary shapes

Aurel BULGAC*
*Department of Physics, University of Washington,
P.O. Box 351560, Seattle, WA 98195–1560, USA*

Piotr MAGIERSKI*
*Department of Physics, University of Washington, Seattle, WA 98195–1560, USA*
*The Royal Institute of Technology, Physics Department Frescati, Frescativägen 24, S–10405, Stockholm, SWEDEN*
*Institute of Physics, Warsaw University of Technology, ul. Koszykowa 75, PL–00662, Warsaw, POLAND*

(April 3, 2018)

## Abstract

A new algorithm for determining the eigenstates of $n$–dimensional billiards is presented. It is based on the application of the Cauchy theorem for the determination of the null space of the boundary overlap matrix. The method is free from the limitations associated with the shape of the billiard and could be applied even for nonconvex geometries where other algorithms face difficulties. Moreover it does not suffer from the existence of eigenvalue degeneracies which is another serious shortcoming of many methods. In the paper we apply the algorithm to a few simple cases where the analytical solutions exist. Numerical solutions have been investigated for the case of annular billiard.

Typeset using REVTEX

# I. INTRODUCTION

In spite of the fact that determining the spectrum of a cavity is certainly more than a century old problem, numerical algorithms do not seem to abound, and moreover, they do not seem always to solve the problem entirely. The references to known methods familiar to us, do not seem to be always the original ones. However, these references apparently cover all the main ideas used so far, with the exception of a few rather specialized algorithms. We shall mention here only those methods which have the potential to be used for a rather large class of billiard shapes and hopefully in more than two dimensions.

The first type of algorithms amounts to determining the roots of a very high order determinant [1–5]. This is not a very pleasant numerical task, as such a determinant easily can vary by many orders of magnitude and the numerically null space is certainly another major problem, which can and apparently does lead to spurious states. The variant suggested in Ref. [5] is somewhat worse in some respect, as it requires determination of complex roots of a complex valued determinant. Even though the spectrum of a cavity is real, when implemented in an obvious and natural way the method used in Ref. [5] leads to spurious states, the nature of which is mysterious. Moreover it has been shown recently that these methods have also problems for nonconvex billiard geometries [6].

In Ref. [7,8] another method was used, which amounts to determining relatively sharp minima of the norm of the wave function over the domain boundary. Such minima do not seem to be always well defined as there are some spurious minima as well, not so deep however. Moreover it is not entirely clear how to disentangle single and multiple degenerate states. The determination of all eigenstates requires multiple runs with randomly chosen conditions [9]. The only way to decide whether all the eigenvalues have been found is by comparing the determined average level density with the Weyl formula. Even though this method seems to reveal all the eigenvalues in the end, it does appear to us to be a rather time consuming procedure, needing a "constant outside intervention and evaluation" of the results and thus requiring too much time.

A significantly superior method, which apparently is free of many of the above drawbacks was recently suggested in Ref. [10]. One of the major deficiencies of this method however is its limitation to the so called star–shaped domains. A star–shaped domain is one for which $\hat{\boldsymbol{n}} \cdot \boldsymbol{r} > 0$, where $\hat{\boldsymbol{n}}$ is the outward normal to the domain boundary and $\boldsymbol{r}$ is the position vector of a point on the domain boundary with respect to an arbitrary origin within the domain. The restriction to star–shaped domains excludes for example annular billiards, peanut–shaped billiards and so forth, which are extremely interesting for various applications. Besides this, a few rather small technical details, concerning its actual numerical implementation have been only slightly covered in Ref. [10] and thus the numerical limitations of this method are not entirely clear. But in any case, this method apparently is the most powerful one available in literature so far.

Another method known to us is the constraint operator method presented in Ref. [11]. Although it allows to determine all eigenvalues in the truncated space, it does not seem to be very accurate for some states, especially the high lying states. The reason could be that the algorithm requires several numerical integrations and diagonalizations. It is also not obvious whether it can actually be used for domains with holes for example.

In this paper we are suggesting an entirely different method, which when correctly imple-mented should lead to the determination of the whole eigenspectrum in a given interval. The method might not necessarily be the fastest, as the method suggested in Ref. [10] is perhaps faster. However, the method presented here should work for arbitrary shapes, unlike the one in Ref. [10]. Its numerical implementation does not seem to suffer from any particular and/or severe drawbacks.

## II. ALGORITHM

The solution of the Helmholtz equation

$$\Delta \phi(\boldsymbol{r}) = -k_0^2 \phi(\boldsymbol{r}) \tag{2.1}$$

in the $n$–dimensional connected domain $\mathcal{D}$ with the Dirichlet boundary conditions on the boundary $\mathcal{B}$

$$\phi(\boldsymbol{r})|_\mathcal{B} = 0 \tag{2.2}$$

can be represented as follows

$$\phi(\boldsymbol{r}) = \sum_{n=1}^{N} c_n \psi_n(k_0, \boldsymbol{r}) \tag{2.3}$$

where functions $\psi_n(k_0, \boldsymbol{r})$ are solutions of the Helmholtz equation (2.1), which nevertheless do not satisfy the boundary condition (2.2). Usually it is the most convenient to choose functions $\psi_n$ in the form of plane waves: $\exp(ik_0\hat{\boldsymbol{k}}_n \cdot \boldsymbol{r})$ with condition: $\hat{\boldsymbol{k}}_n \cdot \hat{\boldsymbol{k}}_n = 1$ (see however Section IIIC). In such a case the sum in (2.3) should be understood as an integral over all possible orientations of the unit vectors $\hat{\boldsymbol{k}}_n$. Since in any numerical realization an integral is always discretized in one way or another, we shall use a discrete summation throughout most of this paper. (The interested reader should find no difficulty in turning this "approximate" representation into an "exact" one, if such a need arises.) The eigenvectors and eigenvalues will be determined from the condition that the boundary norm vanishes, namely that

$$\frac{1}{S} \oint_\mathcal{B} dS w(\boldsymbol{r}) \sum_{n,m=1}^{N} c_n^* \psi_n^*(k_0, \boldsymbol{r}) \psi_m(k_0, \boldsymbol{r}) c_m = \sum_{n,m=1}^{N} c_n^* \mathcal{O}_{nm}(k_0) c_m = 0, \tag{2.4}$$

where we have chosen arbitrarily a unit normalization of the diagonal matrix elements, as $S = \oint_\mathcal{B} dS w(\boldsymbol{r})$. An arbitrary positively defined weight function $w(\boldsymbol{r}) > 0$ can be introduced as well and sometimes this flexibility of the formalism can be profitably used. In this paper we shall limit ourselves to a unit weight function $w(\boldsymbol{r}) = 1$, unless otherwise noted. Obviously, for any finite domain $\mathcal{D}$ the quantization condition Eq. (2.4) is satisfied only for discrete values of $k_0$. The determination of these eigenvalues and of their degeneracies is numerically the most difficult problem. For arbitrary values of $k$ let us introduce the eigenvalues and eigenvectors of the boundary overlap matrix $\mathcal{O}(k)$ (BOM)

$$\mathcal{O}(k) C_\alpha = \lambda_\alpha(k) C_\alpha, \tag{2.5}$$

where $C_\alpha$ is a column vector and the appropriate matrix to vector multiplication is implied. From the non–negativity of the boundary norm it follows that for real $k$ these eigenvalues are non–negative

$$\lambda_\alpha(k) \geq 0. \tag{2.6}$$

Only for an eigenvalue $k_0^2$ of the Helmholtz equation with the Dirichlet boundary conditions one or more BOM eigenvalues vanish, $\lambda_\alpha(k_0) = 0$. The number of such eigenvalues $\lambda_\alpha(k_0)$, which vanish simultaneously, is equal to the degeneracy of the corresponding eigenvalue of the original Helmholtz equation. Even though we did not rigorously proved it yet, it appears that in the neighborhood of an eigenvalue of the Helmholtz equation, those BOM eigenvalues which vanish, always do so quadratically only, i.e.

$$\lambda_\alpha(k) \propto (k - k_0)^2 + \ldots \tag{2.7}$$

The basic idea behind our approach is to analytically continue the BOM $\mathcal{O}(k)$ into the complex $k$–plane, namely to make $k$ complex in the definition of the matrix elements

$$\mathcal{O}_{nm}(k) = \frac{1}{S} \oint_\mathcal{B} dS \psi_n^*(k, \boldsymbol{r}) \psi_m(k, \boldsymbol{r}) \tag{2.8}$$

and to compute around a contour $\mathcal{C}$ in the complex $k$–plane the integral

$$\mathcal{N}(\mathcal{C}) = \oint_\mathcal{C} \frac{dk}{4\pi i} \sum_{\alpha=1}^N \frac{\lambda'_\alpha(k)}{\lambda_\alpha(k)}. \tag{2.9}$$

Here $\lambda'_\alpha(k)$ is the derivative of the eigenvalue $\lambda_\alpha(k)$ with respect to $k$. This derivative can be easily calculated once the eigenvectors of the BOM are known as follows

$$\lambda'_\alpha(k) = C_\alpha(k)^\dagger \mathcal{O}'(k) C_\alpha(k), \tag{2.10}$$

where $\mathcal{O}'(k)$ is the derivative of $\mathcal{O}(k)$ with respect to $k$, $C_\alpha(k)$ are normalized as usual $C_\alpha^\dagger(k) C_\alpha(k) = 1$ and in all these relations obvious vector and matrix multiplications are implied. As an attentive reader would have remarked, we have divided the value of the integral by $4\pi i$, instead of the naively expected $2\pi i$, in order to take into account the double degeneracy of a root of $\lambda_\alpha(k)$, we have discussed above, see Rels. (2.6) and (2.7). Also, when $k$ is complex, when computing the BOM elements the *bra*–vectors are actually functions of $k^*$, while the *ket*–vectors are functions of $k$. Only by defining the BOM elements in this way $\mathcal{O}_{nm}(k)$ is an analytical function, therefore a function of $k$ only and not a function of $k$ and $k^*$ as well. We shall consider here contours $\mathcal{C}$ in the semiplane $\text{Re}(k) > 0$ only. The integral in Rel. (2.9) can be performed analytically, as

$$\mathcal{N}(\mathcal{C}) = \oint_\mathcal{C} \frac{dk}{4\pi i} \sum_{\alpha=1}^N \frac{\lambda'_\alpha(k)}{\lambda_\alpha(k)} = \frac{1}{4\pi i} \sum_{\alpha=1}^N \ln(\lambda_\alpha(k))|_\mathcal{C} \tag{2.11}$$

We do not use this obvious result, as in any computer implementation the determination of the actual Riemannian sheet and of the change of the logarithm around the contour seems to be ambiguous. The numerical evaluation of the integral however, appears to be always straightforward to implement.

$\mathcal{N}(\mathcal{C})$ is thus exactly equal to the number of eigenvalues (counting the number of degeneracies as well), of the Helmholtz equation Eq. (2.1) with the Dirichlet boundary conditions Eq. (2.2), on the segment of the real $k$–axis enclosed by the contour $\mathcal{C}$. Consequently, no eigenvalue of the original equation can thus be missed. This statement is strictly speaking correct only in the limit $N \to \infty$. However, one can easily convince oneself that if $N$ is suitable large, this holds true anyway. Note that $\mathcal{N}(\mathcal{C})$ takes only integer values and thus when $N$ is larger than a certain value the $N \to \infty$ limit is exactly attained (unless the unit vectors $\hat{\boldsymbol{k}}_n$ are distributed in a very peculiar manner of course) (see table I).

One can introduce also other useful quantities. For example

$$\mathcal{S}_n(\mathcal{C}) = \oint_\mathcal{C} \frac{dk}{4\pi i} \sum_{\alpha=1}^N \frac{\lambda'_\alpha(k)}{\lambda_\alpha(k)} k^n = \sum_{\beta \in \mathcal{C}} k_\beta^n, \qquad (2.12)$$

where the sum over $\beta \in \mathcal{C}$ is over all eigenvalues enclosed by the contour $\mathcal{C}$ and $n$ is in principle an arbitrary number, not necessarily an integer and positive (N.B. the origin is not encircled by the contour $\mathcal{C}$). If the contour $\mathcal{C}$ encloses only one eigenvalue then its value is equal to $k_\beta = \mathcal{S}_1(\mathcal{C})$. By determining $\mathcal{S}_n(\mathcal{C})$ for $n = 0, \ldots, \nu$, where $\nu = \mathcal{S}_0(\mathcal{C}) = \mathcal{N}(\mathcal{C})$, one can easily set up a polynomial, whose roots are the eigenvalues enclosed inside the contour $\mathcal{C}$. The functions $\mathcal{S}_n(\mathcal{C})$ have similar convergence properties as $\mathcal{N}(\mathcal{C})$ discussed above. Obviously, one can compute in this way arbitrary functions of the eigenspectrum as well. In particular, it is useful to calculate Fourier components of the quantity $\mathcal{S}_n(\mathcal{C})$:

$$\tilde{\mathcal{S}}_n(\mathcal{C}, t_m) = \oint_\mathcal{C} \frac{dk}{4\pi i} \sum_{\alpha=1}^N \frac{\lambda'_\alpha(k)}{\lambda_\alpha(k)} k^n e^{ikt_m} = \sum_{\beta \in \mathcal{C}} k_\beta^n e^{ik_\beta t_m}, \qquad (2.13)$$

where $t_m = 2\pi \dfrac{m}{\Delta k}, m = -L, ..., L$ and $\Delta k$ is the length of the interval on the real $k-$axis enclosed by the contour $\mathcal{C}$. Thus through the inverse Fourier transform one can obtain e.g. level density $\rho$ or energy distributions $g$ inside the contour $\mathcal{C}$:

$$\rho_\mathcal{C}(k) = \frac{1}{2L+1} \sum_{m=-L}^{L} \tilde{\mathcal{S}}_0(\mathcal{C}, t_m) e^{-ikt_m},$$

$$g_\mathcal{C}(k) = \frac{1}{2L+1} \sum_{m=-L}^{L} \tilde{\mathcal{S}}_2(\mathcal{C}, t_m) e^{-ikt_m}. \qquad (2.14)$$

Here we put $\rho_\mathcal{C}(k) = g_\mathcal{C}(k) = 0$ for $k$ outside the contour. Since the numerical costs of calculations of the Fourier transform $\tilde{\mathcal{S}}_n$ and $\mathcal{S}_n$ are of the same order we can gain at almost no expense more precise information about the distribution of the functions $\mathcal{S}_n(\mathcal{C})$ inside the integration contour.

The practical implementation of this algorithm is rather straightforward. There are essentially a few relatively simple aspects one has to keep in mind:

- One should divide the real $k$–axis in not too long intervals to be enclosed by a complex contour $\mathcal{C}$. When computing the BOM eigenvalues $\lambda_\alpha(k)$ along such a contour one should use always the same number of basis wave functions $\psi_n$. When the basis is increased, even though one might not gain in overall numerical accuracy, there is a side effect. Irrelevant eigenvalues $\lambda_\alpha(k)$ give a large contribution to the integrand in Rel. (2.9) and thus changing the number of wave functions $\psi_n$ along the contour leads to erroneous results.

- The total number of basis wave functions should be chosen somewhat larger than the number of quantum states the boundary can accommodate in this energy range. For example, for a 2–dimensional boundary one should have of the order of $Lk/\pi$ plane waves, where $L$ is the length of the outer perimeter.

- As the method gives the exact number of the eigenstates in a given energy interval, one can easily narrow the interval so as to determine the exact location of any eigenstate and its degeneracy as well, along with the corresponding eigenvector of the Helmholtz equation. It is unavoidable to have parts of the contour $\mathcal{C}$ close to the real $k$–axis. It is profitable however to choose the imaginary part of the contour $\mathcal{C}$ not too far and not to close to the real $k$–axis. A symmetric rectangular contour, with two sides parallel to the real $k$–axis and two sides normal to it seems like a most reasonable and flexible choice. Parts of such a contour can be used repeatedly in order to narrow down the position of the actual eigenvalues.

- The matrix elements of BOM can differ between each other by many orders of magnitude if the basis is large. It can affect the numerical accuracy of the method. Therefore it is recommended in such a case to rescale the BOM matrix: $\tilde{\mathcal{O}}(k) = \mathcal{F}\mathcal{O}(k)\mathcal{F}$, where $det\mathcal{F} \neq 0$.

- The imaginary part of the contour should be chosen at a distance of the order of the mean level separation. At closer distances the integrand changes too rapidly with $k$, while at larger distances from the real $k$–axis a significantly increased and unnecessary numerical accuracy might be required to compute the integrand.

- The fact that $\mathcal{N}(\mathcal{C})$ is integer valued makes its computation somewhat easy, as a relatively low numerical accuracy can be used however.

- The Fourier transform calculated in the finite interval contains spurious components at high frequencies. In such a case we recommend to use a window function (e.g. the Bartlett window) while calculating the inverse transform:

$$\rho_\mathcal{C}(k) = \frac{1}{2L+1} \sum_{m=-L}^{L} \sum_{\beta \in \mathcal{C}} \left(1 - \left|\frac{m}{L}\right|\right) e^{i(k_\beta - k)t_m},$$

$$g_\mathcal{C}(k) = \frac{1}{2L+1} \sum_{m=-L}^{L} \sum_{\beta \in \mathcal{C}} \left(1 - \left|\frac{m}{L}\right|\right) k_\beta^2 e^{i(k_\beta - k)t_m}. \quad (2.15)$$

- If the distance between the eigenvalue and the limit of the interval on the real $k-$axis is smaller than $\dfrac{\Delta k}{2L+1}$, it gives rise to the spurious periodic behavior of the inverse

Fourier transform. To avoid it one should calculate Fourier transformation with enlarged value of $\Delta k$, namely

$$\Delta k \to \Delta k(1+\alpha) \tag{2.16}$$

with a value for $\alpha \approx 0.05\ldots 0.15$.

As we have mentioned above the simplest and perhaps the most useful type of contour $\mathcal{C}$ is a rectangle. Let us define such a rectangle by its four corners $(a - i\delta, b - i\delta, b + i\delta, a + i\delta)$ with $0 < a < b$ and $\delta > 0$. Then the integrals Rel. (2.12) can be somewhat simplified and thus the amount of numerical work reduced:

$$\mathcal{S}_n(\mathcal{C}) = \int_a^b \frac{dx}{2\pi} \sum_{\alpha=1}^N \text{Im}\left[\frac{\lambda'_\alpha(x-i\delta)}{\lambda_\alpha(x-i\delta)}(x-i\delta)^n\right] \tag{2.17}$$
$$+ \int_0^\delta \frac{dy}{2\pi} \sum_{\alpha=1}^N \text{Re}\left[\frac{\lambda'_\alpha(b+iy)}{\lambda_\alpha(b+iy)}(b+iy)^n - \frac{\lambda'_\alpha(a+iy)}{\lambda_\alpha(a+iy)}(a+iy)^n\right].$$

As strange as it might sound, any of the known in the literature methods should encounter most troubles for problems with high symmetries, especially in higher than two dimensions, when high degeneracies are present. In such cases one has to determine not only the presence of an eigenstate, but also its degeneracy. The problems without symmetries are easier in this respect, as degeneracies are rare and accidental. However, once in a while an accidental degeneracy might appear. When tunneling occurs, very closely spaced doublets appear in the spectrum [12]. For the method presented here that does not seem to pose any challenges (except perhaps the actual resolution of such a doublet, but not the determination of its existence). The dimensionality of the problem appears to be largely irrelevant as well, except for the obvious increase of the dimensionality of the BOM. Some of these aspects will be illustrated in the next section.

In many instances one is interested in evaluating the ground state energy of a many fermion system for a given shape and a given number of particles $\mathbf{N}$, that is

$$E_{gs} = \frac{\hbar^2}{2m}\mathcal{S}_2(\mathcal{C}), \tag{2.18}$$

where the contour $\mathcal{C}$ has to be chosen so as to satisfy the condition

$$\mathbf{N} = \mathcal{N}(\mathcal{C}). \tag{2.19}$$

The contour could be a union of several contours, chosen in accordance with the "rules" discussed above. In order to evaluate the ground state energy $E_{gs}$ one needs the position of the Fermi level, which is determined by solving Eq. (2.19). The exact position of the other eigenvalues are not needed and thus one has to determine exactly only a few eigenvalues around the Fermi level.

## III. A FEW PARTICULAR CASES

The examples we are going to discuss here are particularly instructive. In many of these cases some, most or all of the calculations can be performed analytically. Thus these examples serve as an extremely useful guide through the potential numerical problems, some of which we have alluded to in the previous section. In particular, from the exact expressions for the BOM eigenvalues given below, it follows that for $k \to \pm i\infty$ the integrand behaves as $\lambda'_\alpha(k)/\lambda_\alpha(k) \propto 1/k$ and thus it is not profitable to deform the contour $\mathcal{C}$ into a straight parallel to the imaginary $k$–axis, as one might have been tempted to.

### A. 1–dimensional segment

One can easily show that for a 1–dimensional segment of length $L$ one can solve the problem exactly using the outlined method. The BOM has exactly two eigenvalues

$$\lambda_1(k) = 2\sin^2\frac{kL}{2}, \tag{3.1}$$
$$\lambda_2(k) = 2\cos^2\frac{kL}{2}.$$

The contour integrals (2.9) and (2.12) can be evaluated analytically and they lead to the expected exact results. One can see here explicitly that the roots of the BOM eigenvalues have double multiplicities, as we have stated in the previous section.

### B. Circular Billiard

If one considers a circular billiard of radius $R$, the BOM elements can be given analytically

$$\mathcal{O}(\theta, \theta') = J_0\left(2kR\sin\frac{\theta-\theta'}{2}\right). \tag{3.2}$$

Here $J_0(x)$ is the cylindrical Bessel function of first kind and $\theta$ and $\theta'$ determine the direction of two unit vectors in cylindrical coordinates in an obvious manner, $\hat{\boldsymbol{k}} = (\cos\theta, \sin\theta)$. In a rather limited so far numerical implementation of the present method we have established that all of the exact solutions of the circular billiard are reproduced.

One can show that for the circular billiard the exact BOM eigenvectors and BOM eigenvalues (in the continuum limit when the number of basis plane waves is infinite and $N \to \infty$) are $c(\theta) = \exp(im\theta)/\sqrt{2\pi}$ and $\lambda_m = J_m^2(kR)$ respectively, with an arbitrary integer $m$. Therefore, the solution one obtains using the method described in the present paper leads to the exact result for the eigenvectors and eigenvalues of the Helmholtz equation with Dirichlet boundary conditions. Note again the fact that the roots of the BOM eigenvalues $\lambda_m$ have double multiplicities.

## C. Annular Billiard

We shall consider now the billiard of radius $R$ with a circular hard core of radius $a < R$, with an offset $\delta < R - a$ along the $x$–axis (see Fig. 1). For the numerical implementation of the method for the annular billiard it is more convenient to use the basis wave function in the form [13]:

$$\psi_n(kr, \phi_a) = [J_m(kr)Y_m(ka) - Y_m(ka)J_m(kr)] \exp(im\phi_a), \tag{3.3}$$

where $J_m$ and $Y_m$ are cylindrical Bessel functions of the first and second kind, respectively. This implies that the boundary conditions on the inner circle are automatically fulfilled. One can convince oneself that the popular plane wave basis is inadequate in this case. The simplest way to see this is by attempting to use a plane wave basis to represent the present basis wave functions. If the Bessel functions of the first kind can be represented as a superposition of plane wave of a given energy, the Bessel function of the second kind cannot. Since the system is invariant with respect to the $C_2$ symmetry the BOM matrix decomposes into two blocks with the following matrix elements:

$$\mathcal{O}^{(+)}_{mm'} = \frac{1}{2\pi} \int_0^{2\pi} d\phi \cos(m\phi_a)[J_m(kr)Y_m(ka) - Y_m(ka)J_m(kr)] \times$$
$$\cos(m'\phi_a)[J_{m'}(kr)Y_{m'}(ka) - Y_{m'}(ka)J_{m'}(kr)]$$
$$\mathcal{O}^{(-)}_{mm'} = \frac{1}{2\pi} \int_0^{2\pi} d\phi \sin(m\phi_a)[J_m(kr)Y_m(ka) - Y_m(ka)J_m(kr)] \times$$
$$\sin(m'\phi_a)[J_{m'}(kr)Y_{m'}(ka) - Y_{m'}(ka)J_{m'}(kr)], \tag{3.4}$$

where $r = \sqrt{1 + \delta^2 - 2\delta \cos\phi}$, $\cos\phi_a = \dfrac{\cos\phi - \delta}{r}$ and the radius of the billiard is set equal to 1 (see Fig. 1).

The number of basis states needed for the calculations up to the some value $k_{max}$ can be estimated from the condition that the only states contributing to the solution should have at least one node associated with the radial motion inside the domain $\mathcal{D}$ i.e. $\exists_{r \in (a, 1+\delta)} |\psi_{n \leq N}(k_{max}, r)|^2 = 0$. An increase of the basis affects the computation time. The main contribution comes from the diagonalization of the BOM matrix along the integration contour. In the table I we have shown the dependence of the computation time on the number of basis states $\psi_n$. Although the computation time is sensitive also to the precision of the integration both in the $(r, \phi)$– and $k$–spaces, it depends approximately linearly on the number of integration points whereas it exhibits the quadratic behavior as a function of the number of basis states.

In the Figure 2 we have shown results of calculations performed for $a = 0.5$ and $\delta = 0, 0.25$. Since for the 2-dimensional system the density of states is in the first approximation independent of $k^2$ we have used the variable $\epsilon = k^2$ instead of $k$. The level density $\rho$ was obtained using $L = 8$ Fourier components in each interval $\Delta\epsilon = 60$, see Eq. (2.14). The number of particles and the energy are expressed in the form:

$$N = \int_0^\mu d\epsilon\, \rho(\epsilon) = \int_0^\mu d\epsilon \sum_i \rho_{\mathcal{C}_i}(\epsilon)$$
$$E = \int_0^\mu d\epsilon\, g(\epsilon) = \int_0^\mu d\epsilon \sum_i g_{\mathcal{C}_i}(\epsilon), \tag{3.5}$$

where $\mu$ is the chemical potential. Since we are rather interested in the fluctuating part of these quantities we have substracted the smooth behavior associated with the density $\rho_0$ calculated from the Weyl formula:

$$\rho_0(\epsilon) = \frac{1-a^2}{4} - \frac{1+a}{4\sqrt{\epsilon}}. \tag{3.6}$$

Then the corresponding particle and energy fluctuations are given by:

$$\delta N(\epsilon) = N(\epsilon) - \frac{1-a^2}{4}\epsilon - \frac{1+a}{\sqrt{\epsilon}}$$
$$\delta E(N) = E(N) - \int_0^{\mu_0} d\epsilon \rho_0(\epsilon)\epsilon, \tag{3.7}$$
$$\tag{3.8}$$

where $\mu_0$ is determined by the condition:

$$N = \int_0^{\mu_0} d\epsilon \rho_0(\epsilon). \tag{3.9}$$

In Ref. [14]. it was shown that the annular billiard is chaotic for $\delta > 0$. However the degree of chaoticity i.e. the fraction of the phase space occupied by chaotic trajectories depends on the eccentricity parameter $\delta$. The system becomes fully chaotic for $\delta = 1 - a$ when all the trajectories must hit the inner circle. Thus the shell effects visible in the Fig. 2 for $\delta = 0.25$ are remnants of the ordered motion still existing in this case, see Refs. [15]. Associated orbits are characterized by the impact parameter $L = \sin\alpha > a + \delta$ (see Ref. [14]) and forever encircle the inner disk. One can observe that the period of oscillations of $\delta E$ is larger for $\delta = 0.25$. It is caused by the fact that the shortest periodic orbits giving rise to the shell effects at $\delta = 0$ i.e. triangular and rectangular orbits, are destroyed in the $\delta = 0.25$ case.

Within the method presented in this paper we are also able to plot the density distribution of particular states. In Fig. 3 we have presented the densities of two states for $\delta = 0.0$ and $\delta = 0.25$ belonging to two representations of $C_2$ denoted by $(+)$ and $(-)$.

### D. $n$–dimensional billiards

For billiards in any dimensions the BOM elements have simple forms if either the entire boundary or only parts of it have spherical/ellipsoidal (or 2$d$–circular/ellipse) shape, planar/linear and/or any combinations of these shapes. The sphere/circle could be easily deformed into an ellipsoid as well and one would still get simple expressions for the BOM elements. In particular for a spherical 3$d$–cavity one can show, similarly to the case of the 2$d$–circular cavity, that the BOM eigenstates are as expected the spherical harmonics, i.e. $c_{lm}(\theta, \phi) = Y_{lm}(\theta, \phi)$, and the BOM eigenvalues are $\lambda_{lm} = j_l^2(kR)$ for $m = -l, ..., l$, with spherical Bessel functions instead of cylindrical ones. This again shows that the BOM method leads to the correct eigenstates.

## IV. CONCLUSIONS

The method introduced here appears flexible enough to allow essentially a foolproof calculation of the entire spectrum of an arbitrarily shaped billiard in any dimensions. We believe also that the method can be optimized as well so as to speed up numerical calculations. In particular, the introduction of evanescent plane waves [16] is one aspect which is worth a serious consideration, as it apparently leads to significant increase of the numerical accuracy [10]. Other boundary conditions, besides Dirichlet, can be implemented in a straightforward manner too. We were in way rather surprised to realize that the present method (which of course is nothing but a consequence of Cauchy's theorem) is not apparently being used in numerical analysis, as its implementation is rather straightforward. The fact that this method is a "foolproof" one should make it a standard one whenever finding roots of a rather large class of one variable functions is somewhat of a challenge.

## ACKNOWLEDGEMENTS

We thank C.H. Lewenkopf for providing several literature leads, G.F. Bertsch for suggesting the 1–dimensional example and DOE for partial financial support. One of the authors (PM) thanks the Nuclear Theory Group in the Department of Physics at the University of Washington for hospitality and acknowledge the financial support from the Swedish Institute and the Göran Gustafsson Foundation.

# REFERENCES

\*     E–mail: bulgac@phys.washington.edu

\*     E–mail: piotr@msiw49.msi.se, Piotr.Magierski@fuw.edu.pl

TABLES

TABLE I. Computation time estimates for the annular billiard with $\delta = 0.25$. The contour $\mathcal{C}$ has been choosen to enclose the interval $(960, 1030)$ on the real $\epsilon$ axis.

| No. of basis states | Time (sec.) | $\mathcal{N}(\mathcal{C})$ |
|---|---|---|
| 2  | 11.13  | 0.000  |
| 4  | 15.38  | 0.000  |
| 6  | 21.87  | 0.000  |
| 8  | 49.49  | 0.000  |
| 10 | 58.08  | 0.000  |
| 12 | 67.44  | 1.001  |
| 14 | 79.39  | 1.000  |
| 16 | 90.82  | 2.001  |
| 18 | 106.78 | 2.000  |
| 20 | 121.18 | 1.999  |
| 22 | 140.28 | 4.015  |
| 24 | 159.56 | 5.000  |
| 26 | 185.45 | 7.000  |
| 28 | 206.25 | 7.001  |
| 30 | 235.27 | 9.000  |
| 32 | 258.33 | 9.001  |
| 34 | 300.64 | 11.000 |
| 36 | 319.14 | 11.000 |
| 38 | 361.64 | 11.000 |
| 40 | 393.90 | 11.000 |

FIGURES

FIG. 1. Annular billiard.

FIG. 2. The density of states ($\rho$), particle number fluctuations ($N - N_{smooth}$) and energy fluctuations ($E - E_{smooth}$) as a function of $\epsilon = k^2$ for annular billiard.

FIG. 3. Selected densities of quantum states of the annular billiard (the sign in paranthenses refers to the $C_2$ symmetry of the corresponding wave function): a) $k = 26.1, \delta = 0$, (+); b) $k = 26.1, \delta = 0$, (−); c) $k = 25.354, \delta = 0.25$, (+); d) $k = 25.096, \delta = 0.25$, (−).

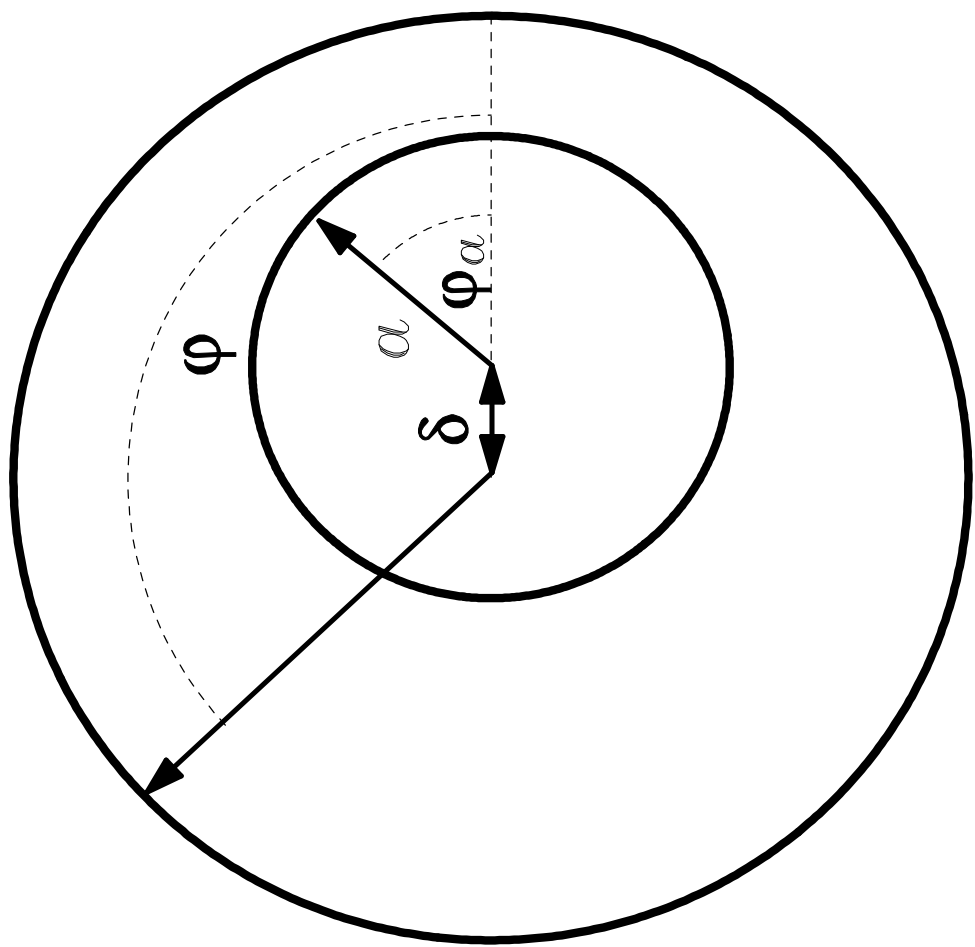

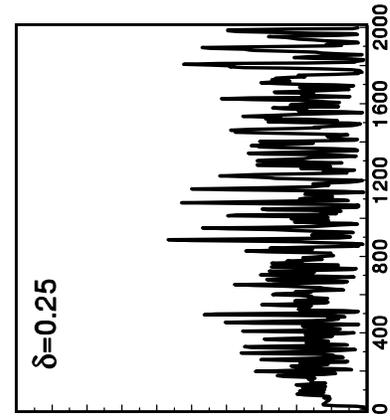
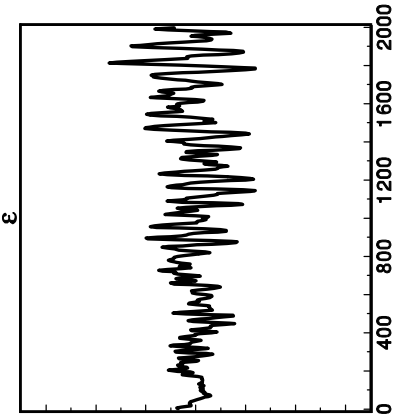
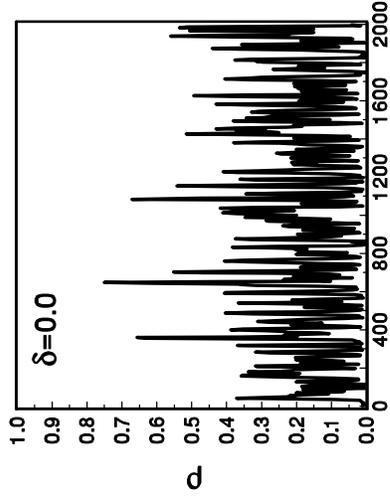
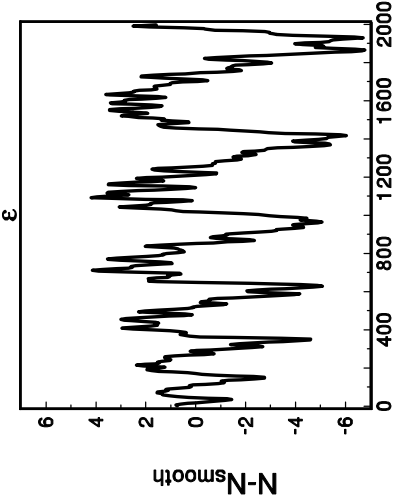

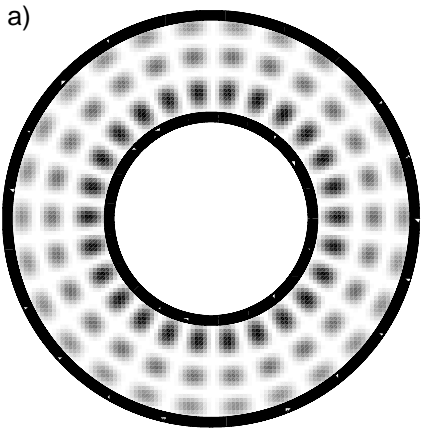
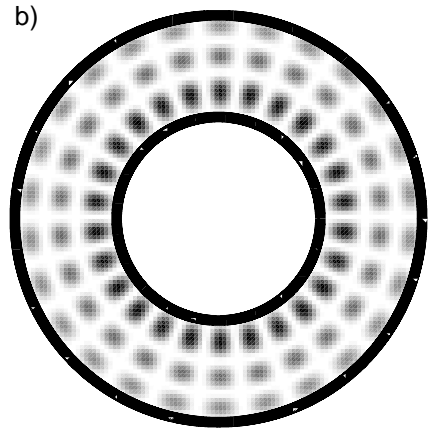
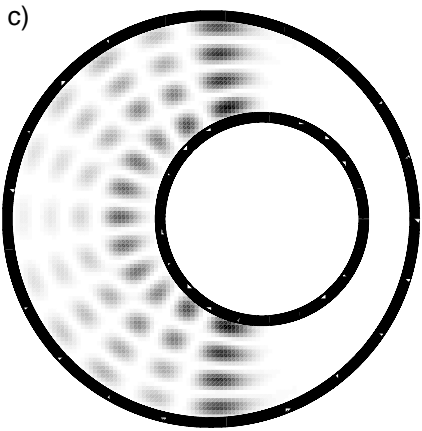
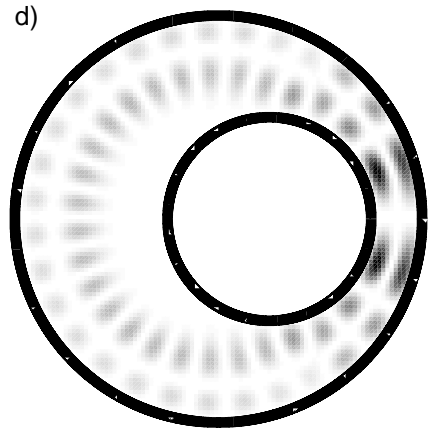